\documentclass[twocolumn,prb,superscriptaddress,longbibliography]{revtex4-2}

\usepackage{amsmath,amsfonts,amssymb}
\usepackage{graphicx,color}
\usepackage{hyperref}
\usepackage{epstopdf}
\usepackage{float}
\usepackage{multirow}
\usepackage{hhline}
\usepackage{ulem}
\usepackage{mathtools}
\usepackage[caption=false]{subfig}
\usepackage{soul}
\usepackage[english]{babel}

\usepackage{graphicx}
\usepackage{dcolumn}
\usepackage{bm}

\usepackage[utf8]{inputenc} 

\usepackage{multirow}
\usepackage[warn]{mathtext}
\usepackage{amssymb}

\usepackage{textcomp} 
\usepackage{indentfirst} 
\usepackage{amsmath} 
\usepackage{graphicx}
\DeclareGraphicsExtensions{.pdf,.png,.jpg}
\usepackage{pgfplots}
\pgfplotsset{compat=1.13}

\usepackage{hyperref}
\hypersetup{
    colorlinks=true,
    linkcolor=blue,
    filecolor=black,      
    urlcolor=blue,
    citecolor= violet
}

\usepackage{algpseudocode}

\usepackage{adjustbox}
\usepackage{tabularx}

\usepackage[nottoc]{tocbibind}
\usepackage{mathtools}

\usepackage{xcolor}
\definecolor{C0}{HTML}{4C72B0}
\definecolor{C1}{HTML}{DD8452}
\definecolor{C2}{HTML}{55A868}
\definecolor{C3}{HTML}{C44E52}

\begin{document}

\makeatletter
\newcommand{\settitle}{\maketitle}
\makeatother

\title{Low-overhead quantum error correction codes with a cyclic topology}

\author{Ilya A. Simakov}
\email{simakov.ia@phystech.edu}
\affiliation{National University of Science and Technology ``MISIS'', 119049 Moscow, Russia}
\affiliation{Russian Quantum Center, 143025 Skolkovo, Moscow, Russia}
\affiliation{Moscow Institute of Physics and Technology, 141701 Dolgoprudny, Russia}
\author{Ilya S. Besedin}
\thanks{Present address: Department of Physics, ETH Zurich, Zurich, Switzerland}
\affiliation{National University of Science and Technology ``MISIS'', 119049 Moscow, Russia}
\affiliation{Russian Quantum Center, 143025 Skolkovo, Moscow, Russia}

\date{\today}

\begin{abstract}
Quantum error correction is an important ingredient for scalable quantum computing. 
Stabilizer codes are one of the most promising and straightforward ways to correct quantum errors, are convenient for logical operations, and improve performance with increasing the number of qubits involved.  
Here, we propose a resource-efficient scaling of a five-qubit perfect code with increasing-weight cyclic stabilizers for small distances on the ring architecture, which takes into account the topological features of the superconducting platform.
We show an approach to construct the quantum circuit of a correction code with ancillas entangled with non-neighboring data qubits.
Furthermore, we introduce a neural network-based decoding algorithm supported by an improved lookup table decoder and provide a numerical simulation of the proposed code, which demonstrates the exponential suppression of the logical error rate.

\end{abstract}

\settitle

\section{Introduction}

Large error rates of about $10^{-3}$ achieved on qubit computers \cite{Arute2019, chen2021exponentia, google2023suppressing, Krinner_2022,  PhysRevLett.129.030501, PhysRevLett.127.180501} are a major obstacle towards practical application of quantum computing. Therefore, quantum error correction (QEC) is expected to be an essential and unavoidable part of a fault-tolerant quantum computer \cite{PhysRevA.32.3266, PhysRevA.52.R2493, Calderbank_1996, knill1996threshold, Kitaev_2003, RevModPhys.87.307}. One of the most promising methods for correcting quantum errors is considered to be stabilizer codes, in which logical qubits are encoded in an entangled state of a large number of physical qubits. By measuring $X$ or $Z$ parity operators (syndromes) one can identify the occurrence of errors in the underlying physical qubits \cite{DiVincenzoShor, SteaneQEC, GottesmanPhd}. A variety of such codes have been developed and simulated under plausible error conditions \cite{LowDistanceSurfaceCodes, Trout_2018, Muyuan2017, XZZXSurfaceCode, Bravyi2018, OBrien2017, Hastings_2021, Gidney2021faulttolerant, Gidney_2022, takada2024improvingthresholdfaulttolerantcolor, domokos2024characterizationerrorscnotsurface, AWScatcode, Yalecatcode}.

Experimental implementation of correction codes is a difficult task, requiring a large number of qubits and high-precision operations on them. At the same time, especially in large devices, the topology of the code should be mapped to the underlying qubits and their interconnections to keep the number of operations per code size low. For example, in superconducting systems, high-fidelity two-qubit gates can typically be realized only between adjacent qubits. When choosing a correction code, this aspect should be taken into account. Among recent experimental demonstrations, we highlight the repetition code \cite{chen2021exponentia}, the surface code \cite{google2023suppressing, Krinner_2022, PhysRevLett.129.030501, DiCarlo2021logical, andersen2020repeated, acharya2024quantumerrorcorrectionsurface}, and the color code \cite{Bluvstein_2023, PhysRevX.11.041058, ryananderson2022implementingfaulttolerantentanglinggates} implemented on different platforms.

Another important property of the correction codes is scalability, which means the ability to correct more physical errors by increasing the number of data qubits involved. The milestone experiments demonstrating error suppression are presented in papers \cite{chen2021exponentia, google2023suppressing}. 

The smallest possible code that corrects an arbitrary single-qubit error is the five-qubit perfect code \cite{PhysRevLett.77.198}. Its stabilizers can be generated as a cyclic permutation of a base stabilizer $XZZXI$. The five-qubit code implementation with iSWAP gates on a ring topology was proposed in \cite{schuch2003natural} and thoroughly simulated in \cite{chain10, Antipov_2022, du2024fault}. There is also the cyclic analogue of the toric code \cite{PhysRevA.84.062319, Xu_2022}, which can be considered as a scaling of the five-qubit code.

In this paper we describe a set of low-distance QEC codes where the stabilizers are generated by cyclic permutations of the basic stabilizer, similar to the five-qubit code \cite{Grassl_2000}. 
The main advantage of this code is that the number of required physical qubits grows linearly with the code distance, although this is accompanied by an increase of the stabilizer weights.
For practical implementation, we propose a quantum circuit consisting of alternating ancillary and data qubits forming a ring, with natural two-qubit iSWAP and SWAP gates available between any neighboring qubits. 
The proposed code requires reasonably low hardware overhead and is affordable in near-term demonstration experiments.
We also present a neural network decoder that preprocesses the prediction results of the lookup table decoder (LUT) together with the raw measurements of the ancillary qubits.
We compare the performance of this decoder to a simple LUT for a simple error model, and demonstrate the same asymptotic scaling with a lower pre-factor.
Finally, we briefly discuss the limitation of the code performance under a standard depolarizing circuit-level noise model and outline possible further improvements of the circuit, addressing the hardware-inspired errors.

\section{Code construction}

\begin{figure*}[t]
    \center{\includegraphics[width=\linewidth]{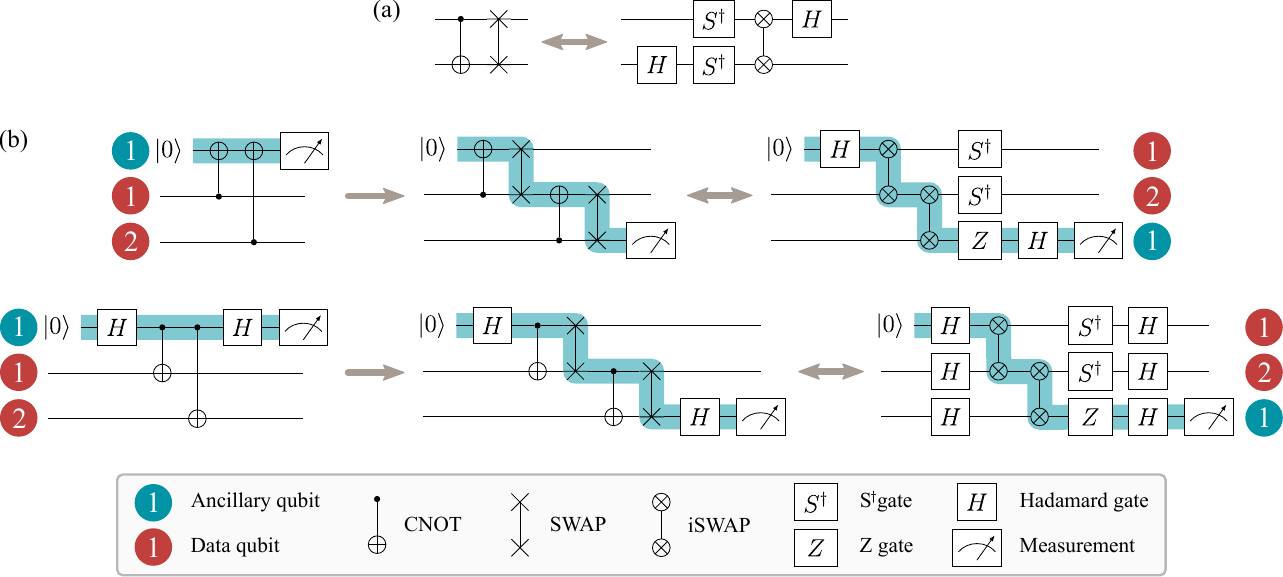}}
    \caption{(a) Implementation of two consecutive CNOT and SWAP gates with a single iSWAP operation and single qubit rotations. 
    (b) Transformation of the conventional quantum circuit with $ZZ$ (top) and $XX$ (bottom) stabilizer measurements to a scheme with iSWAP gates and a ``sliding'' ancillary qubit.}
    \label{fig:syndromeconstruction}
\end{figure*}

A quantum error correction code encoding a single logical qubit into $n$ physical qubits can be described by $n-1$ syndrome generators. 
Below we give the basic syndrome for the linear cyclic codes with distances $3,$ $5,$ $7$, and $9$, respectively:
\begin{equation}
    \begin{aligned}
        &d=3, \; n=5 : \; g_0 = ZXXZI \\
        &d=5, \; n=13 : \; g_0 = ZIXXIZ\underbrace{I...I}_{7} \\
        &d=7, \; n=21 : \;  g_0 = ZIIXXXXIIZ\underbrace{I...I}_{11} \\
        &d=9, \; n=29 : \;  g_0 = ZXXIIXIIXIIXXZ\underbrace{I...I}_{15}, 
    \end{aligned}
    \label{eq:stabilizers}
\end{equation}
where $X$, $Y$, $Z$, $I$ are the Pauli operators, and we assume tensor product sign $\otimes$ between each of them. 
The remaining $n-2$ generators can be obtained from $g_0$ by shifting the qubit indices, modulo $n$. 
The logical operators of the codes can be defined as product of single-qubit operators:
\begin{equation}
    X_L = X^{\otimes n}, \; Y_L = Y^{\otimes n}, \; Z_L = Z^{\otimes n}.
    \label{eq:logical_operators}
\end{equation} 

To build an efficient quantum circuit based on the native for superconducting devices two-qubit iSWAP gates \cite{schuch2003natural, PhysRevApplied.10.054062, PhysRevA.96.062323, PhysRevResearch.2.033447, PRXQuantum.5.020338} we first introduce the concept of ``sliding'' ancillary qubits. 
Consider an extraction of the basic $ZZ$ and $XX$ syndromes. 
The standard quantum circuits for such operations are shown on the left panels of Fig.~\ref{fig:syndromeconstruction}(b). 
They contain a single ancillary qubit, which is highlighted in teal.
Through two CNOT gates, it entangles with two data qubits, depicted in red.

\begin{figure*}[t]
    \center{\includegraphics[width=\linewidth]{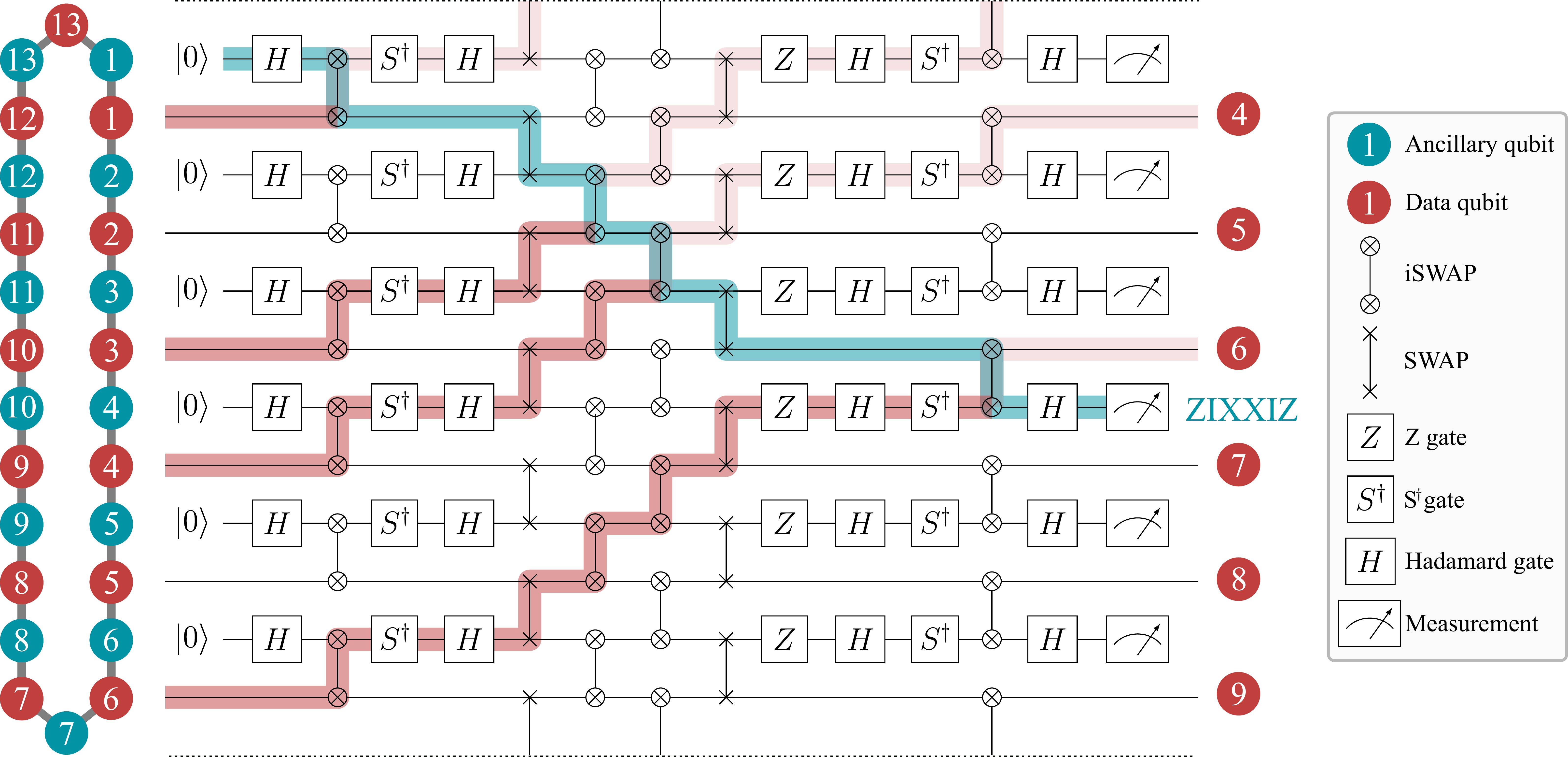}}
    \caption{Schematic layout of the 13-qubit code on the ring topology and a part of the quantum circuit. The red and aquamarine circles mark data and ancillary qubits, respectively. The complete quantum circuit can be reconstructed by translating of the shown segment. The main part of the scheme are the six series of the two-qubit gates applied to the adjacent qubits. We highlight the path of the computational state of the first ancillary qubit and show the data qubits with which it interacts during the correction cycle. We also emphasize that the computational states of the data qubits are shifted at the end of the cycle.}
    \label{fig:chain26}
\end{figure*}

When dealing with superconducting quantum devices, where the two-qubit gates are implemented only between adjacent qubits, one should take into account the topological features of the available processor.
Thereby, it is required to move the ancillary qubit next to the data qubit before performing the entangling operation.
The straightforward way to implement this movement is to utilize SWAP gates, but such an approach increases the circuit depth and correspondingly decreases the code performance.
Fortunately, the sequential CNOT and SWAP operations can be realized by a single iSWAP gate. 
The transformation of conventional $ZZ$ and $XX$ stabilizer measurements to circuits with a ``sliding'' ancillary qubit is shown in Fig.~\ref{fig:syndromeconstruction}(a, b).
This approach enables quantum circuits where it is essential for a qubit to interact with several non-neighboring qubits.

Leveraging the concept of ``sliding'' information qubits, syndrome measurement can be efficiently realized on a ring of alternating $n$ ancillary and $n$ data qubits. 
Assuming that single-qubit gates, initialization, and single-shot readout are available for all qubits, we propose a quantum circuit for the QEC cycle with only iSWAP and SWAP two-qubit gates between the neighboring qubits. 
We note that the scheme contains an extra ancilla, which corresponds to a syndrome that is not independent, but a product of the other syndrome generators.

The code with $d=3$ is well-known as the perfect five-qubit code.
Its behavior under a realistic noise model in the ring topology of 10 serially connected superconducting qubits has been studied in detail in \cite{chain10}. 
Here we focus on the higher distance codes and explain their main concept using the example of the 13-qubit $d=5$ code. 
The quantum circuit of the error correction cycle is shown in Fig.~\ref{fig:chain26}. 
It requires 26 physical qubits, half of which maintain the encoded logical state, and the rest are used to extract syndrome data, which provides information about possible errors that may have occurred during previous cycles. 
Each correction cycle begins and ends with a ground-state initialization and a single-shot readout of all ancillary qubits. 
During the correction cycle, each ancilla undergoes six two-qubit gates and as a result entangles with four data qubits, as highlighted for the first ancilla in Fig.~\ref{fig:chain26}. 
In such a circuit, the ancillary and data qubits move in opposite directions along the ring of physical qubits.
The quantum circuits of the other codes with larger distances are constructed analogously.

Next, we determine the resource requirements for the proposed cyclic code circuits. 
The number of physical qubits involved for the construction of the code of distance $d \in \{3, 5, 7, 9\}$ is $N = 8d - 14$. 
Another hardware resource demanding parameter is the two-qubit gates, which in superconducting devices are generally implemented by external control of a coupling element. 
The basic entanglement operation of the discussed circuits is an iSWAP. 
The SWAP gate can be realized with three iSWAP or three $\sqrt{\text{iSWAP}}$ gates \cite{schuch2003natural}, or even with a controllable $\text{fSim}$ operation, if available \cite{PhysRevLett.125.120504}.
Therefore, the SWAP gate does not require any additional elements on the chip design, and the number of distinct two-qubit gates between physical qubits in the cyclic code is also $N_\text{2Q}=8d - 14$. 
Thus, the proposed cyclic code scales linearly with the required resources, defined as the sum of the total number of physical qubits and the number of different two-qubit gates.
Additionally, in Table~\ref{tab:total_gates}, we present the total number of quantum operations required for a single correction cycle of the proposed code.

\begin{table}[b]
    \centering    
    \begin{tabularx}{\columnwidth}{@{}  *5{>{\centering\arraybackslash}X}@{}}
    \hline
    \hline
         Distance & 3 & 5 &  7 & 9 \\
         \hline
         iSWAP & 20 & 52 & 126 & 232 \\
         SWAP & 0 & 26 & 84 & 174 \\
         $H$ & 20 & 52 & 84 & 116 \\
    \hline
    \hline
    \end{tabularx}
    \caption{Total amount of quantum operations required for a single correction cycle of the proposed code.}
    \label{tab:total_gates}
\end{table}

\begin{figure}[t]
    \center{\includegraphics[width=\linewidth]{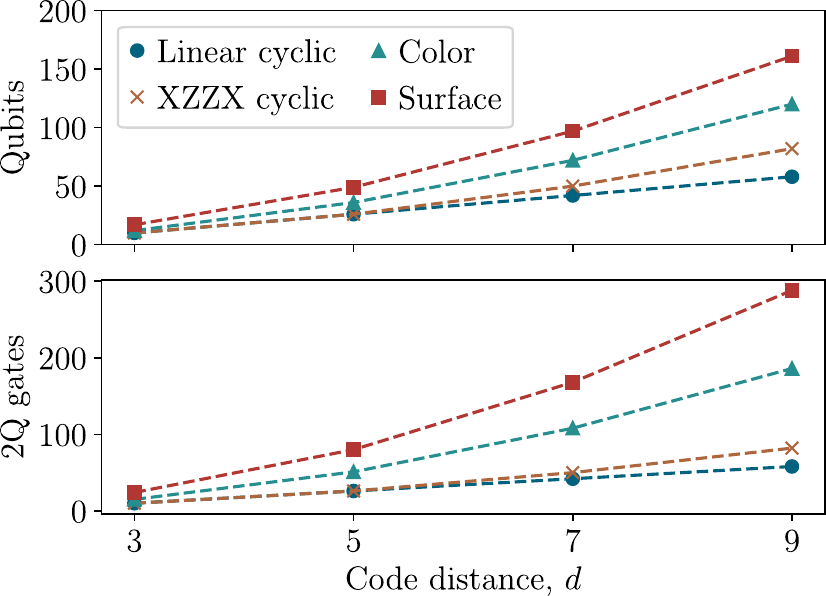}}
    \caption{Comparison of the resource requirements of the proposed code with some promising quantum error correction codes. The concept of resources refers to the total number of physical qubits involved (top) and two-qubit gates implemented on different qubit pairs (bottom). The technical requirements of the proposed cyclic code (blue circles) scale linearly, while the requirements of the cyclic toric (orange crosses), triangular color (green triangles), and rotated surface (red squares) codes scale quadratically. The dashed lines show the trend of each dependence.}
    \label{fig:resources}
\end{figure}

\begin{figure*}[t]
    \center{\includegraphics[width=\linewidth]{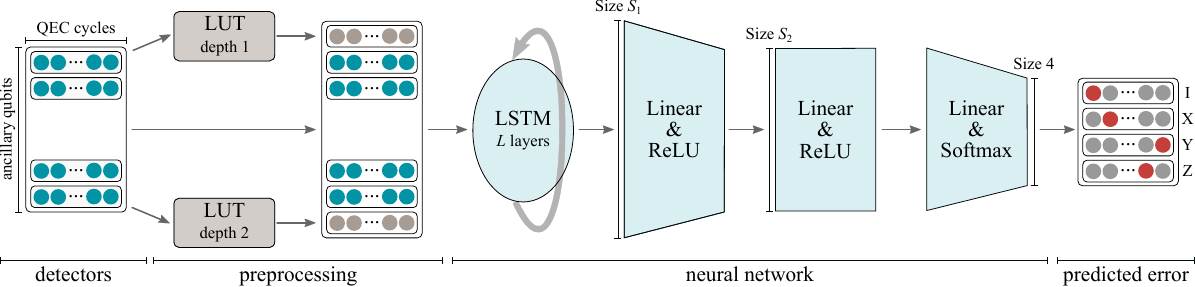}}
    \caption{Decoding algorithm based on the recurrent neural network. First, we produce detection events from raw ancillary qubit measurements during the quantum memory experiment. Then we process the detectors by two lookup table decoders with memory depth 1 and 2 and finally feed the data to the neural network. As a result one gets the probabilities of the occurred errors and selects the most likely correction.}
    \label{fig:nn}
\end{figure*}

In addition, we compare the resource requirements of the proposed cyclic code with popular surface \cite{PhysRevA.86.032324}, color \cite{PhysRevLett.97.180501} and cyclic toric \cite{PhysRevA.84.062319, Xu_2022} codes, see Fig.~\ref{fig:resources}.
To implement a surface code of distance $d$ one needs $N = 2d^2-1$ physical qubits and $N_\text{2Q} = 4d(d-1)$ two-qubit gates.
The triangular color code circuit contains $N=(3 d^2 - 1)/2$ qubits and $N_\text{2Q}= 3(7d^2 - 8d + 1)/8$ different two-qubit gates.
The cyclic toric code is defined by a register shift of a basic syndrome $ZI^{\otimes(t-1)}XXI^{\otimes(t-1)}ZI^{\otimes{(n-2t-2)}}$, where $t=(d-1)/2$.
It can be implemented on a similar cyclic architecture as the proposed code with iSWAP and SWAP gates, and in a such case and requires $N = N_\text{2Q} = d^2 + 1$ qubits and 2Q gates.
We remark that for distances $3$ and $5$ this code and the proposed one are exactly the same.
The difference starts at distance $7$.
We note that the amount of two-qubit gate operations required for syndrome extraction is actually higher than that of other codes, due to the higher weights of the syndromes.

\section{Results}

To characterize the efficiency of the correction codes, we focus on the performance of the logical qubit as a quantum memory. 
In order to simulate a state preservation experiment, we need to exploit a relevant error model and propose a decoding algorithm.

\subsection{Error model}

Here we consider to use the so-called phenomenological error model \cite{Dennis_2002, PhysRevA.109.052438, Srivastava_2022}.
In such an approximation there are two different types of errors.
The first is the depolarization error, which affects the data qubits between QEC cycles while the ancillary qubits are being measured.
Mathematically, it is described by an independent and identically distributed standard depolarization channel:
\begin{equation}
    \mathcal{E}(\rho) = (1-p)\rho + p_x X\rho X + p_y Y\rho Y + p_z Z\rho Z,
\end{equation}
where $p=p_x+p_y+p_z$ is the error probability, and $p_x$, $p_y$, and $p_z$ are probabilities of $X$, $Y$, and $Z$ errors, respectively.
The second error type is a measurement error known as timelike. 
It occurs when the bit-flip error channel $\mathcal{E}_\text{bit}(\rho) = (1-q)\rho + q X\rho X$ with probability $q$ acts on the ancillary qubit just before the Z-basis measurements. 
Such errors do not affect the logical state directly, but they corrupt the measured syndromes and complicate the decoding problem considerably. 
In practice, we use the Clifford stabilizer approach \cite{Aaronson2004} implemented  in the Stim package \cite{gidney2021stim} to simulate the proposed quantum circuits. 
We choose a symmetric depolarization channel: $p_x=p_y=p_z=p/3$ and also take the time-like error probability $q$ equal to $p$.

\subsection{Decoding algorithm}

Building a reliable and fast decoder is a formidable problem of QEC codes. 
For surface codes, the typical approach is graph matching. Assuming that independent errors trigger at most two syndrome elements, decoding can be represented as a graph matching problem, where each node corresponds to a syndrome, and errors correspond to edges. Within this picture, the most likely error sequence can be obtained using the MWPM algorithm \cite{Dennis_2002}. Higher-weight errors, \textit{i.e.}, errors that trigger more than two syndromes at the same time, can be decoded using more sophisticated algorithms, such as belief matching \cite{Higgott2023}. However, this method also involves creating a matching graph that is not suitable for mixed $X$ and $Z$ errors with high-weight syndromes. Again, decoding with a standard belief propagation algorithm is computationally hard for high-distance codes \cite{poulin2008iterativedecodingsparsequantum}.

An alternative approach to address the problem is utilizing neural networks \cite{Krastanov_2017, PhysRevLett.119.030501, Breuckmann2018scalableneural, Chamberland_2018, PhysRevA.99.052351, Overwater_2022, Krastanov_2017, PhysRevLett.122.200501, Gicev2023scalablefast}. The goal of the decoding process is to analyze periodic measurements of ancillary qubits and predict logical errors. Essentially, we need to recognize patterns in numerical sequences, a task well-suited for artificial neural networks. Given that we collect time-series-like data during QEC experiments, recurrent neural networks (RNNs) emerge as strong candidates for the decoding procedure. In this work, we employ the long short-term memory (LSTM) architecture \cite{lstm}, a subtype of RNN that has demonstrated high performance in recent research \cite{Baireuther2018machinelearning, Baireuther_2019}.

The distinguishing feature of our algorithm is that, in addition to syndrome changes (detectors), we provide as an input to the neural network the corrections obtained from a lookup table (LUT) decoder. This approach resembles feature engineering in standard machine learning. While neural networks are powerful for data analysis, they struggle with complex, noisy data, such as that encountered in QEC experiments, often requiring large datasets, substantial GPU power, and processing time. Our method simplifies and accelerates the training process by utilizing predictions from other decoding algorithms. Additionally, we can feed the neural network predictions from multiple decoders, each offering distinct advantages.

We utilize a modified LUT decoder as a supporting mechanism. A basic lookup table stores a mapping between logical errors and the syndromes generated by errors on data qubits. This approach effectively manages spacelike errors but fails with timelike errors. To mitigate these issues, we introduce the concept of memory depth. The idea is straightforward: when decoding the $k$-th correction cycle, if the $k$-th syndrome is trivial, we assume no error has occurred and move to the next cycle. If there are detection events in the $k$-th cycle, we examine the following $D$ cycles. If no non-trivial detectors appear in these $D$ cycles, we correct the error in the $k$-th cycle according to the lookup table. Conversely, if there are non-trivial detectors, we add the $k$-th cycle's syndrome to the detectors of the $(k+1)$-th cycle, deferring correction in the $k$-th cycle.

We refer to the parameter $D$ as the depth of memory. For $D=0$, there is no memory, which corresponds to a basic LUT decoder; if $D=1$, the decoder addresses measurement errors; and if $D=2$, the decoder manages situations when two consecutive measurement errors cancel the syndrome. A significant drawback of this method is that accumulating syndromes also adds spacelike errors, potentially reducing the effectiveness of the lookup table. However, the advantage of this approach is that it produces several fast decoders capable of handling various error scenarios, simplifying neural network decision-making. We denote the memory depth of the LUT decoder as $D$ in brackets, e.g. LUT(1).

Mathematically, the concept of the LUT decoder memory depth \( D \) can be framed as a data preprocessing technique. We define the syndrome changes $s^i_j$ as follows:
\begin{equation}
    s^i_j= 
    \begin{cases}
    0, & \text{if } j = 1\\
    (m^i_{j} - m^i_{j-1}) \bmod 2, & \text{otherwise,}
    \end{cases}
\end{equation}
where \( m^i_j \) represents the result of the ancillary qubit measurements for \( i=1,...,n \) after cycle \( j=1,...,k \), taking values of \( 0 \) or \( 1 \). Consequently, the algorithm for syndrome preprocessing can be written as follows: 
\begin{algorithmic}
    \For {$j \leq m$}
        \If {$\sum_i s^i_j \neq 0 \ \mathbf{and} \ \sum_{k=1}^{D} \sum_i s^i_{j+k} \neq 0$} 
            \For {$i \leq n$}
                \State $s^i_{j+1} \gets \left( s^i_j + s^i_{j+1} \right) \bmod 2$
                \State $s^i_{j} \gets 0$
            \EndFor
        \EndIf
    \EndFor
\end{algorithmic}

Returning to the decoding problem, we input raw stabilizer measurements results into the neural network alongside predictions from two LUT decoders with memory depths $D = \{0, 1\}$. 
The neural network architecture comprises $L$ LSTM layers, each with a hidden size of $ S_1 $, followed by two linear layers of sizes $ S_1 \rightarrow S_2 $ and $ S_2 \rightarrow S_2 $ with a ReLU activation function. 
The final linear transformation dimension is $ S_2 \rightarrow 4 $. A softmax function is used to convert the outputs into a valid probability of four different Pauli operator corrections to the logical qubit state: $X$, $Y$, $Z$, or $I$ (indicating no error). 
We compute the loss function as the negative log-likelihood between the predicted error and the measured logical qubit state, as defined in equation (\ref{eq:logical_operators}) \cite{chain10}. 
The complete decoding pipeline, along with the detailed neural network architecture, is illustrated in Fig.~\ref{fig:nn}. 
In the discussion section, we argue for the efficiency of the preprocessing procedure on the performance of the neural network-based decoder.

\subsection{Code performance}

To evaluate the performance of the code, we simulate a quantum memory experiment for the proposed code with different single-qubit error probabilities $p$.
For a training dataset, we collect $1.5 \times 10^6$ trajectories of a 20-cycle state preservation experiment with probability $p=0.01$ and a logical qubit initialized in three different bases $Z$, $X$, and $Y$. 
The neural network hyperparameters $L$, $S_1$ $S_2$ obtained by a grid search for each code distance are given in Table \ref{tab:nn_parameters}.
To test the code performance we repeat the quantum memory experiment with a logical qubit initialized in three different bases $Z$, $X$, and $Y$ up to 50 correction cycles for $3 \times 10^6$ times.
Finally, we fit the obtained data with the function \cite{OBrien2017, chen2021exponentia}
\begin{equation}
    F(n) = \frac{1}{2} + \frac{1}{2} (1 - 2\epsilon)^{n-n_0},
    \label{fid_approx}
\end{equation}
where $n$ is the number of correction cycles, $\epsilon$ is the logical error rate per correction cycle and the correction fidelity $F(n)$ is defined as the fraction of correctly recovered states. 
The parameters $\epsilon$ and $n_0$ are computed by the least squares method.

\begin{figure}[t]
    \center{\includegraphics[width=\linewidth]{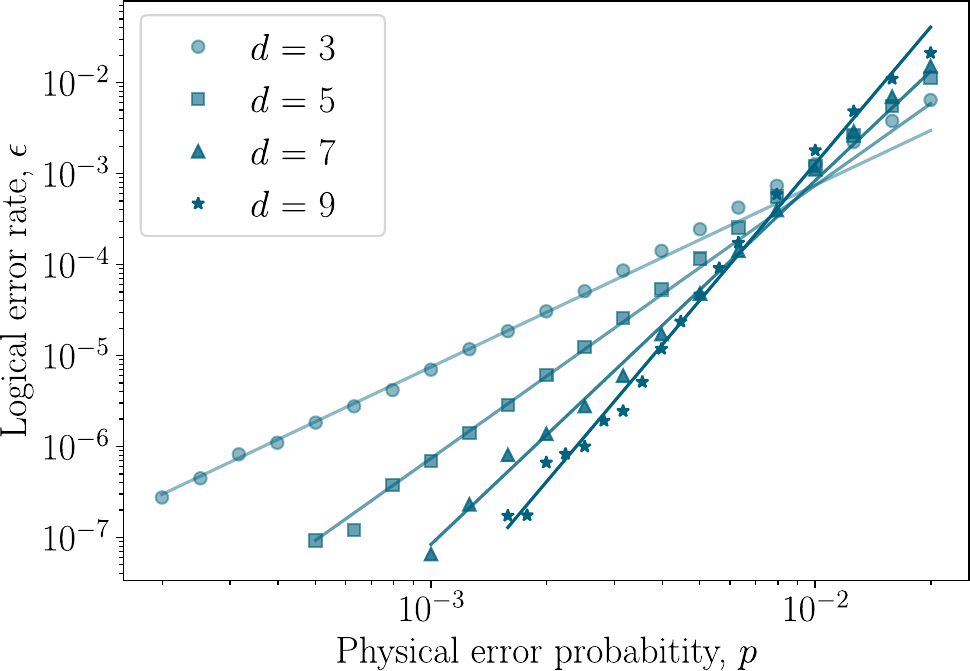}}
    \caption{Logical error rate of the proposed cyclic code plotted as a function of the physical error probability $p$. The lines show the fit of the simulated data with logical error below $10^{-4}$ with the formula~(\ref{eq:power_law}). }
    \label{fig:result}
\end{figure}

Ultimately, we obtain the logical failure rate $\epsilon$ for different distances of the proposed code as a function of the error probability $p$ and depict the corresponding plot in Fig.~\ref{fig:result}. 
Since a QEC code with distance $d$ detects up to $(d-1)/2$ independent single-qubit errors, the relationship between these parameters is expected to be described by a power law: 
\begin{equation}
    \epsilon \sim p^{(d+1)/2}.
    \label{eq:power_law}
\end{equation}
Using a least-square fit with the free parameter $p$, we obtain an estimate for the effective code distance with respect to the simulated error model.
The slope of the curve matches the expected value of $d$, confirming that the code demonstrates the declared code distance within the simulated error model and decoder.
Thus, for the chosen error model, the proposed cyclic code demonstrates the required property of the declared distance.

\begin{table}[b]
    \centering    
    \begin{tabularx}{\columnwidth}{@{} *4{>{\centering\arraybackslash}X}@{}}
    \hline
    \hline
         Code distance & LSTM layers & Size $S_1$ & Size $S_2$ \\
         \hline
         3 & 2 & 256 & 128 \\
         5 & 3 & 256 & 128 \\
         7 & 4 & 256 & 128 \\
         9 & 5 & 512 & 256 \\
    \hline
    \hline
    \end{tabularx}
    \caption{Neural network parameters used in the current work.}
    \label{tab:nn_parameters}
\end{table}

\begin{figure}[t]
    \center{\includegraphics[width=\linewidth]{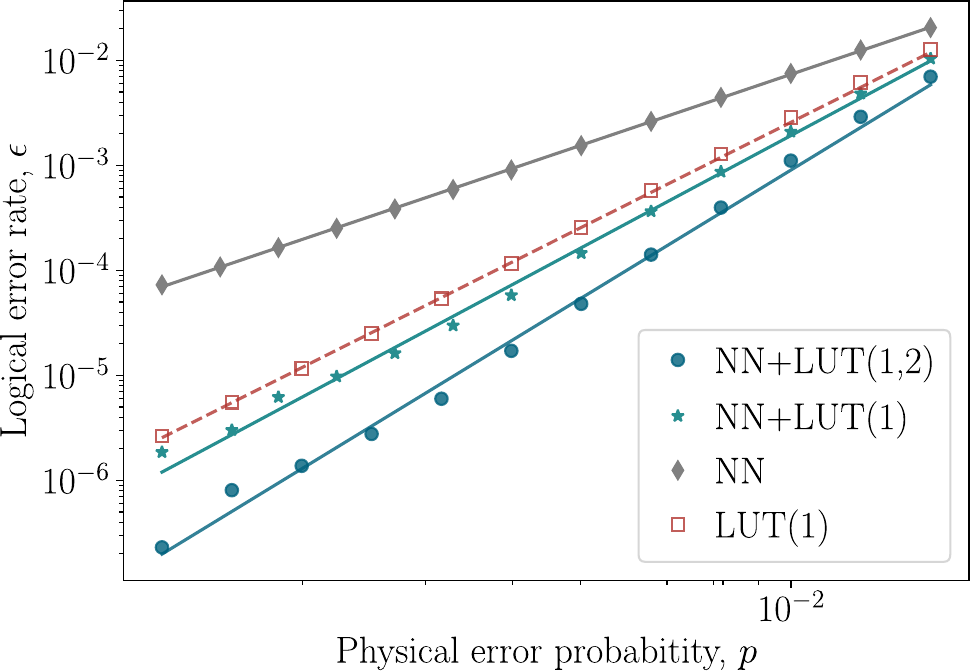}}
    \caption{Logical error rate of a distance-7 cyclic code as a function of the physical error probability $p$ for different decoding algorithms. The filled circles, stars, and diamonds correspond to the results obtained with a neural network trained on detectors and two LUTs with memory depth $D = \{1, 2\}$ predictions, detectors and LUT outcome of depth 1, and bare detectors only. The empty square represents the best performance of a LUT decoder. The lines show the linear fit of the simulated data in logarithmic scale.}
    \label{fig:dec_efficiency}
\end{figure}

\section{Discussion}

Here, we compare the performance of the neural network decoder to the conventional LUT decoder, as well as our modified LUT decoder with memory.
To demonstrate the benefit of the procedure, we focus on the decoding of the cyclic code of distance 7.
The logical error rates as a function of the physical error probability $p$ for different decoding methods are shown in Fig.~\ref{fig:dec_efficiency}. 
The filled circles, stars, and diamonds represent the results obtained with a neural network trained on detectors and two LUTs with memory depth $D = \{1, 2\}$ predictions, detectors and LUT outcome of depth 1, and bare detectors only, respectively. 
For comparison, we also include the best result obtained with the LUT(1) decoder, indicated by empty red squares. 
Notably, the neural network that does not utilize additional data shows a higher logical error rate than the LUT.
Conversely, the neural network combined with two different LUT decoders achieves the best results, surpassing the performance of the neural network with a single LUT decoder.

Yet another important aspect to address is the performance of the proposed codes under the standard circuit depolarization error model. 
We conducted simulations for the 5-qubit and 13-qubit codes and observed effective distances of 1 and 3, respectively. 
Notably, for the 5-qubit code, a full-system density matrix simulation demonstrated an effective distance of 3 \cite{chain10}. 
The key difference between these two models lies in the two-qubit depolarization errors introduced by two-qubit gates, which cause the presence of hook errors \cite{Dennis_2002}. 
A promising approach to address hook errors is adding special flag qubits into the circuit \cite{Chamberland2018flagfaulttolerant, PhysRevLett.121.050502, PRXQuantum.1.010302}. 
As a direction for future research, we believe that introducing flag qubits could significantly enhance cyclic circuits and improve the overall performance under hardware-inspired noise models.

\section{Conclusion}

In conclusion, we propose a resource-efficient scaling of the five-qubit perfect code for small distances within a realistic superconducting circuit topology. 
The corresponding quantum circuit is defined on a ring of superconducting qubits, utilizing two-qubit iSWAP and SWAP gates available between each pair of adjacent qubits.
The total number of qubits required for the code approaches the Hamming bound. 
Today, when state-of-the-art superconducting quantum processors typically consist of several dozens of physical qubits, the small footprint of the correction code presents a significant advantage that allows us to explore QEC experimentally. 
However, we note that the code is not scalable in the conventional sense. 
The maximum number of non-identity single-qubit operators, or in other words the weight of the code syndromes, increases with the code distance, which limits the applicability of these codes for higher code distances. 

We also introduce a decoding algorithm based on the LSTM neural network, which main feature is the dual preprocessing of the raw data by a lookup table decoder with different settings.
Furthermore, we have studied the logical fidelity of the proposed cyclic code assuming a phenomenological error model. 
Despite the inseparability of $X$ and $Z$ syndromes and an increased syndrome weight, we observe an exponential reduction of the logical error rate with code distance using the neural network decoder.
The key challenges for further advancing this code are the improvement of the cyclic circuit with the addition of flag qubits to deal with hook errors and the experimental investigation of the dominant error channels associated with the iSWAP operation.
Addressing these factors will be crucial for enhancing the robustness and applicability of the code and the proposed decoding algorithm in practical quantum computing.

\section*{Acknowledgments}

The authors are grateful to Alexey Ustinov for helpful discussions and critical comments on the manuscript. 
We acknowledge the support of the Russian Science Foundation (Grant No. 21-72-30026) and the Ministry of Science and Higher Education of the Russian Federation in the framework of the Program of Strategic Academic Leadership “Priority 2030” (Strategic Project Quantum Internet).

\renewcommand{\bibname}{Reference}
\normalem{}
\bibliography{main}

\end{document}